\documentclass[a4paper,11pt]{article}
\usepackage{pos}

\usepackage{color}
\usepackage{caption}

\title{$\Lambda$ and $\Sigma$ potentials in neutron stars, hypernuclei, and heavy-ion collisions
}

\author*[a]{Asanosuke Jinno}
\author[b]{Koichi Murase}
\author[c]{Yasushi Nara}

\affiliation[a]{Department of Physics, Faculty of Science, Kyoto University,
 Kyoto, 606-8502, Japan}

\affiliation[b]{Department of Physics, Tokyo Metropolitan University,
Hachioji 192-0397, Japan}

\affiliation[c]{Akita International University,
 Yuwa, Akita-city 010-1292, Japan}

\emailAdd{jinno@ruby.scphys.kyoto-u.ac.jp}
\emailAdd{murase@tmu.ac.jp}
\emailAdd{nara@aiu.ac.jp}

\abstract{
	With an appropriate $YNN$ force,
	the $\Lambda$ single-particle potential ($\Lambda$ potential) can be made strongly
	repulsive at high density, and one can solve the hyperon puzzle of neutron stars.
	We investigate the consistency of such a $\Lambda$ potential, evaluated recently
	from $YN$ and $YNN$ forces based on chiral effective field theory,
	with hypernuclear data and heavy-ion collision data.
	It is found that model calculations with such a $\Lambda$ potential can reproduce
	the data of the $\Lambda$ hypernuclear spectroscopy and the $\Lambda$ directed flow in
	heavy-ion collisions. Also, we evaluate the $\Sigma$ potential, which can be
	calculated by using the same hyperon forces	as for the $\Lambda$ potential.
	Specifically, we show that the low-energy constants characterizing
	the strength of the $YNN$ force can be chosen to suppress the appearance of
	the $\Lambda$'s in neutron stars while at the same time
	the empirical value of the $\Sigma$ potential is reproduced.
}

\FullConference{International Conference on Exotic Atoms and Related Topics and Conference on Low Energy Antiprotons (EXA-LEAP2024)\\
 26-30 August 2024\\
Austrian Academy of Sciences, Vienna.\\}

\begin{document}
\maketitle

\section{Introduction}
\label{sec:Intro}

The hyperon puzzle of neutron stars refers to the observation that most equations
of state (EOSs) with hyperons are too soft to support the observed massive neutron
stars~\cite{Demorest:2010bx}. One promising solution is that the three-baryon
forces (3BFs) among hyperon ($Y$) and nucleons ($N$)
are strongly repulsive so
that the hyperons cannot appear in neutron stars.
For example, in Ref.~\cite{Gerstung:2020ktv},
$YNN$ forces have been constructed based on the decuplet dominance
approximation~\cite{Petschauer:2016pbn} and employed together with a $YN$ potential
derived within chiral effective field theory ($\chi$EFT)~\cite{Haidenbauer:2013oca}.
With such a combination, a $\Lambda$ single-particle potential ($\Lambda$ potential)
fulfilling that scenario can be obtained.

In this contribution, we examine whether the $\Lambda$ potentials published
in Ref.~\cite{Gerstung:2020ktv} are consistent with the $\Lambda$ separation energies
of hypernuclei and with the directed flow $v_1$ of $\Lambda$ in heavy-ion collisions.
Furthermore, we investigate the impact of the $YNN$ interactions proposed in Ref.~\cite{Gerstung:2020ktv}
on the $\Sigma$ single-particle potential ($\Sigma$ potential).

Let us emphasize that our work is in line with a recent trend in nuclear
matter studies~\cite{Drischler:2021kxf,Huth:2021bsp,Rutherford:2024srk},
conducting a unified approach integrating nuclear experiments and neutron star
observations with the modern nuclear force from $\chi$EFT to obtain a
well-constrained EOS of dense matter.  This approach
contributes to a more comprehensive understanding of both nuclear experiments
and astrophysical observations.
For a microscopic description of EOS, the properties of hyperons in nuclear matter should also be constrained.
We utilize the experimental data involving hyperons to evaluate the validity of
the existing $\Lambda$ potential, which is important in determining the onset density
of strangeness.

\section{Evaluating the repulsive $\Lambda$ potential from $\Lambda$ hypernuclear data}
\label{sec:Hypernuclei}

\begin{figure}[hbp]
	\centering
	\begin{minipage}[b]{.4\textwidth}
	\includegraphics[width=0.9\hsize]{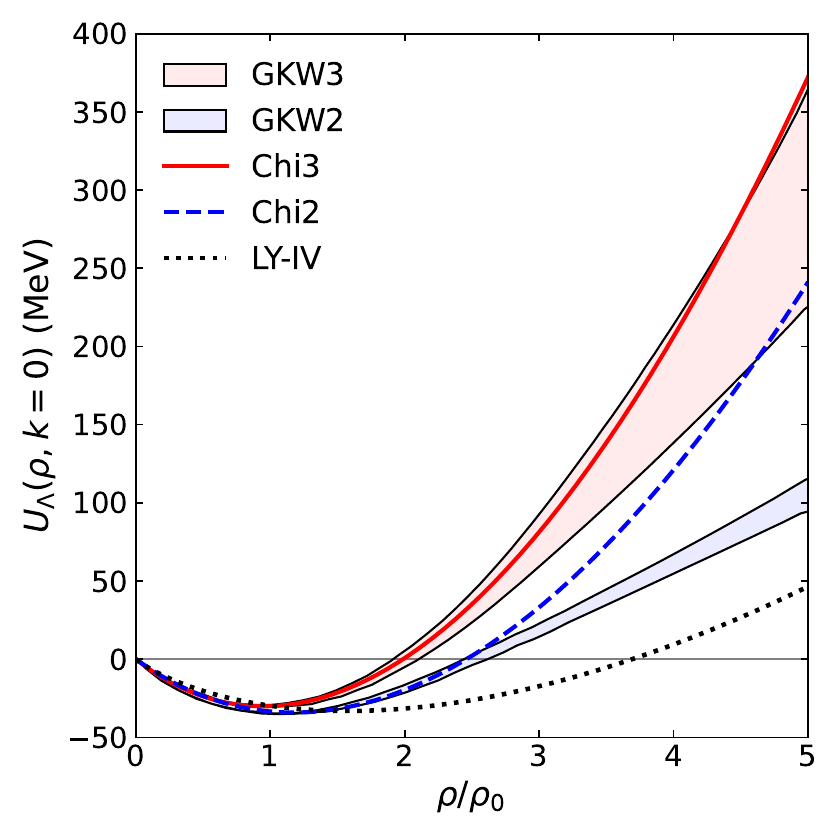}
	\end{minipage}
	\begin{minipage}[b]{.5\textwidth}
	\includegraphics[width=0.9\hsize]{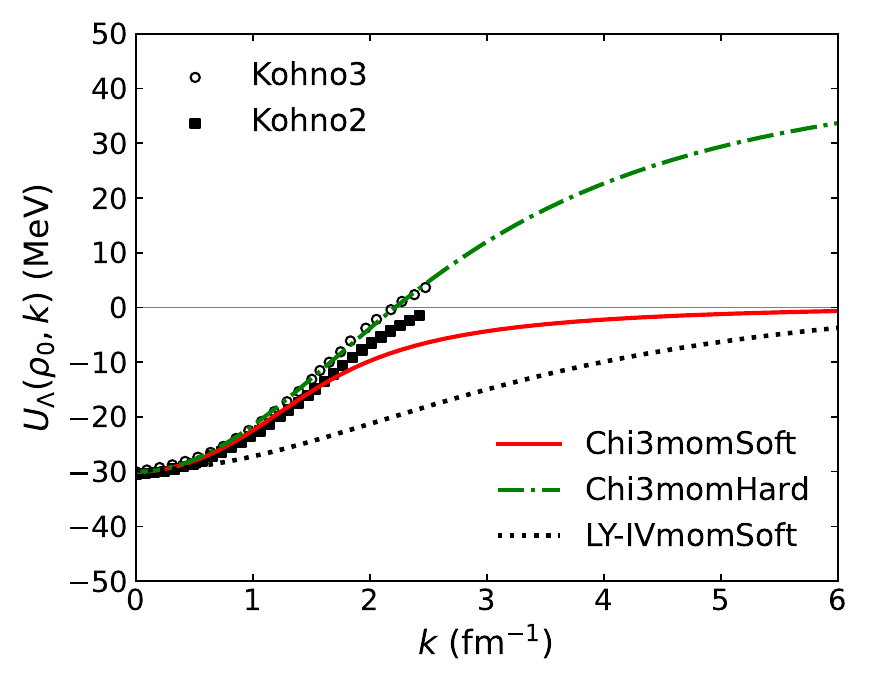}
	\end{minipage}
	\caption{(left panel) Density dependence of the $\Lambda$ potential.  GKW3
	represents the results from $\chi$EFT with $YN$ and $YNN$
	forces~\cite{Gerstung:2020ktv}.  GKW2 is also from $\chi$EFT but without the
	3BFs~\cite{Gerstung:2020ktv}.  Chi3 (solid line) and Chi2 (dashed
	line) are fitted to GKW2 and GKW3 up to $\rho/\rho_0<1.5$, respectively.
	LY-IV (dotted line) is a conventional $\Lambda$ potential~\cite{Lanskoy:1997xq}.
	(right panel) Momentum dependence of the
	$\Lambda$ potential.  Kohno3 represents the result from $\chi$EFT with $YN$ and $YNN$
	forces~\cite{Kohno:2018gby}.  Kohno2 is the result from
	$\chi$EFT without the 3BFs~\cite{Kohno:2018gby}.  Chi3momSoft
	(solid line) and Chi3momHard (dash-dotted line) are constructed to reproduce
	Kohno3 up to $2.5~\text{fm}^{-1}$ and $1.0~\text{fm}^{-1}$, respectively.
	LY-IVmomSoft (dotted line) is fitted to the momentum dependence of LY-IV up
	to $1.0~\text{fm}^{-1}$.}
	\label{fig:potentials}
\end{figure}

First, we utilize the $\Lambda$ hypernuclear spectroscopy to examine the $\Lambda$ potential.
We consider three $\Lambda$-potential models (Chi3, Chi2, and LY-IV) as follows
(see Ref.~\cite{Jinno:2023xjr} for more details):
We constructed the Chi3 potential by fitting the result of
$\chi$EFT with $YN$ and $YNN$ forces~\cite{Gerstung:2020ktv,Kohno:2018gby}
to the Skyrme-type $\Lambda$ potential~\cite{Jinno:2023xjr}.
The $YN$ force is chosen as NLO13(500)~\cite{Haidenbauer:2013oca},
while the $YNN$ force is constructed by the decuplet dominance
approximation~\cite{Petschauer:2016pbn}. For reference, the Chi2 potential was
similarly constructed without the 3BF\@.  The LY-IV potential is a
conventional $\Lambda$ potential~\cite{Lanskoy:1997xq} attractive at high
density, with which $\Lambda$'s appear in dense neutron star matter.
The density dependence of the $\Lambda$ potentials is plotted on the left panel
of figure~\ref{fig:potentials}.
The momentum dependencies for
Chi2, Chi3, and LY-IV in the lower momentum region $k\le 1.0~\text{fm}^{-1}$
exhibit behaviors
similar to those of Kohno2, Kohno3, and LY-IVmomSoft shown on the right panel of figure~\ref{fig:potentials}, respectively
(see Ref.~\cite{Jinno:2023xjr} for the comparison).
We employ the Skyrme-Hartee-Fock method using the above-mentioned three
different $\Lambda$ potentials.
One parameter that cannot be determined from
the uniform-matter results is tuned to reproduce the $\Lambda$ binding energy
data of $^{13}_{~\Lambda} \mathrm{C}$\@.

\begin{figure}[hbp]
\begin{minipage}[h]{.5\textwidth}
\includegraphics[width=0.9\hsize]{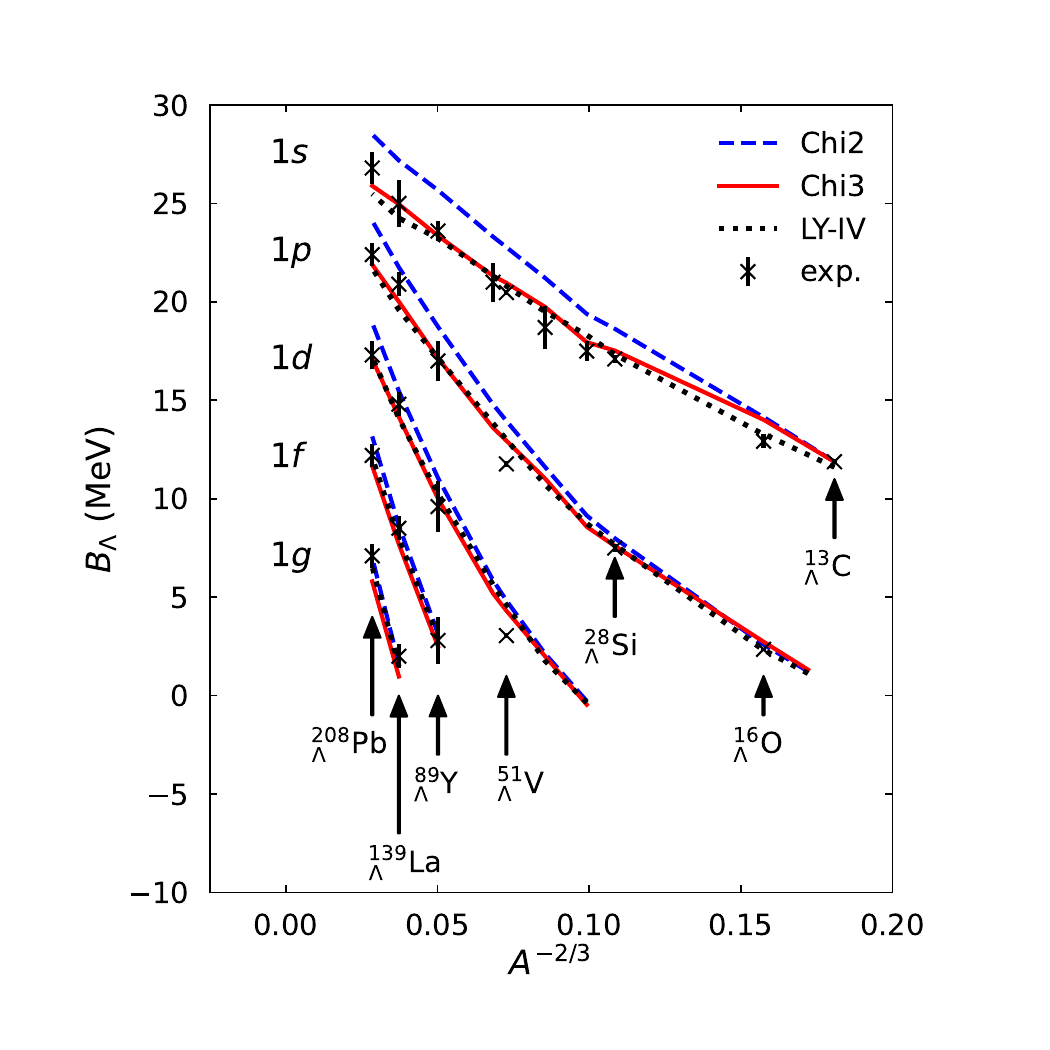}
\end{minipage}
\begin{minipage}[h]{.5\textwidth}
\includegraphics[width=0.9\hsize]{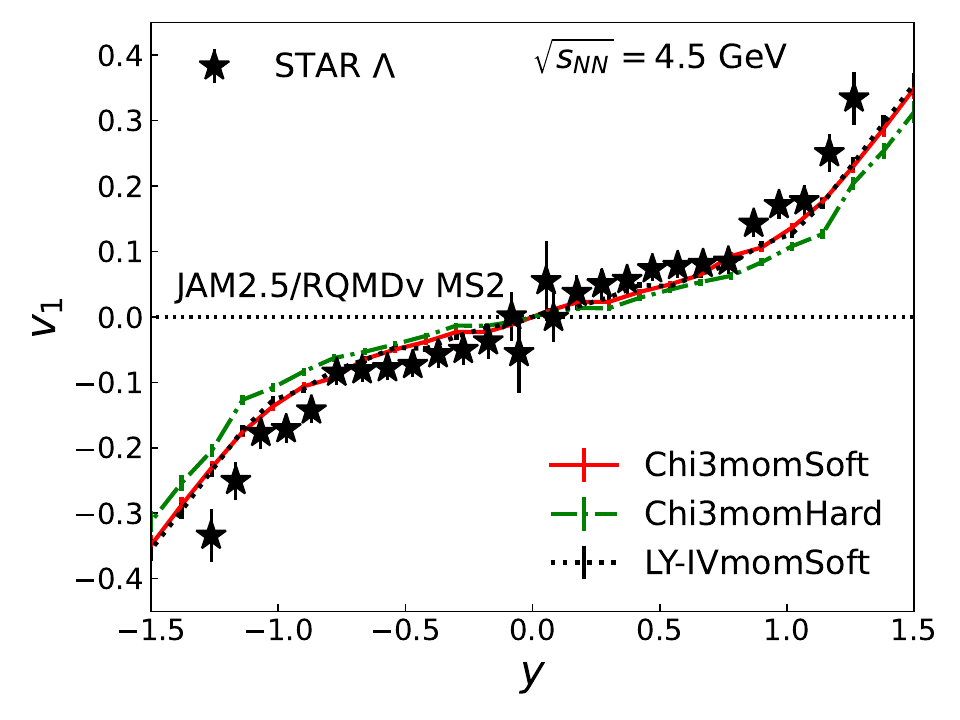}
\end{minipage}
\caption{(left panel) $\Lambda$ binding energy of the $\Lambda$
hypernuclei. Experimental data (cross) can be found in
Ref.~\cite{Jinno:2023xjr}. The figure is adopted from Ref.~\cite{Jinno:2023xjr}.
(right panel) Directed flow of $\Lambda$ in mid-central Au+Au collisions at
$\sqrt{s_{NN}}=4.5~\text{GeV}$. The STAR data are taken from
Ref.~\cite{STAR:2020dav}.
The figure is updated from Ref.~\cite{Nara:2022kbb} by using the updated version of \texttt{JAM2}.
}
\label{fig:results}
\end{figure}

We compare the model calculations with data on the separation energies of $\Lambda$
hypernuclei~\cite{Jinno:2023xjr} on the left panel of figure~\ref{fig:results}.
Chi3 reproduces the data as accurately as LY-IV\@.
In contrast, we found that Chi2 overbounds by several $\text{MeV}$ due to the
excessive potential depth at the saturation density $\rho_0 \approx 0.16~\text{fm}^{-3}$.
Thus, Chi2 can be excluded, yet we need other data to
constrain the repulsion of the $\Lambda$ potential at high densities.

\section{Evaluating the repulsive $\Lambda$ potential from heavy-ion collision data}
\label{sec:heavy-ion collision}

Next, we consider the rapidity dependence of the $\Lambda$ directed flow
in heavy-ion collisions~\cite{Nara:2022kbb},
\begin{equation}
v_1=\langle \cos \phi\rangle
=\Biggl\langle \frac{p_x}{\sqrt{p_x^2+p_y^2}}\Biggr\rangle,
\end{equation}
where $\phi$ is the azimuthal
angle measured from the reaction plane and $p_x$ and $p_y$ are the transverse
momenta of a particle. We use the Lorentz vector version of the relativistic
quantum molecular dynamics (RQMDv) model~\cite{Nara:2021fuu} implemented in the
\texttt{JAM2} transport
code\footnote{\url{https://gitlab.com/transportmodel/jam2}}.

In the heavy-ion collision simulation, the high momentum part of the momentum
dependence is important. We construct Chi3momSoft, Chi3momHard, and LY-IVmomSoft by extrapolating
the momentum dependence of Chi3 and LY-IV to a high momentum region by assuming the Lorentzian form:
\begin{align}
    \label{eq:Um_HIC}
    U_m(\rho(x),k) = \dfrac{C}{\rho_0} \int d^3 k' \dfrac{f(x,k')}{1+\left[(\boldsymbol{k}-\boldsymbol{k}')/\mu\right]^2},
\end{align}
where $C$ and $\mu$ are fitting parameters
and $f(x,k)$ is the single-particle
distribution function.  In the actual heavy-ion simulations, we implement the
momentum-dependent potential as the Lorentz vector
$U^\mu_m$~\cite{Nara:2022kbb}.  Since $\chi$EFT is not entirely reliable above
the momentum cutoff of $550~\text{MeV} \simeq
2.8~\text{fm}^{-1}$~\cite{Kohno:2018gby}, we prepared two variations:
Chi3momHard and Chi3momSoft are constructed to reproduce the $\chi$EFT
result~\cite{Kohno:2018gby} up to $2.5~\text{fm}^{-1}$ and
$1.0~\text{fm}^{-1}$, respectively.  LY-IVmomSoft is constructed to reproduce
the momentum dependence of LY-IV with Eq.~\eqref{eq:Um_HIC} up to
$1.0~\text{fm}^{-1}$.  The momentum dependence of the $\Lambda$ potentials is
plotted on the right panel of figure~\ref{fig:potentials}.  We note that the
density dependence of Chi3momSoft and Chi3momHard is almost identical to that
of Chi3, as is LY-IVmomSoft to LY-IV.

The results of $v_1$ of $\Lambda$ in mid-central Au + Au collisions at
$\sqrt{s_{NN}}=4.5~\text{GeV}$ are shown on the right panel of
figure~\ref{fig:results} and compared with the STAR data~\cite{STAR:2020dav}.
One can see that both Chi3momSoft and LY-IV reproduce $v_1$ of $\Lambda$ with
equal accuracy, which implies that $v_1$ of $\Lambda$ is not so sensitive to the
density dependence of the $\Lambda$ potential.  On the other hand, Chi3momHard
underestimates $v_1$ of $\Lambda$, which indicates that $v_1$ of $\Lambda$ is
sensitive to the momentum dependence of the $\Lambda$ potential.  Experimental
information on the optical potential of $\Lambda$ may be useful for reducing
the model uncertainty.

\section{How about the $\Sigma$ potential?}
\label{sec:Sigma}

As already mentioned above, in Ref.~\cite{Gerstung:2020ktv}, the 3BFs have been
adjusted in such a way that the $\Lambda$ potential is sufficiently repulsive
at high density so that the appearance of $\Lambda$ hyperons in neutron stars
is suppressed. This is possible for different combinations of the low-energy constants
(LECs),
$H_1$ and $H_2$, that characterize the strength of the 3BF (see the solid lines
in figure 6 of Ref.~\cite{Gerstung:2020ktv}). For other combinations, cf. the
dashed lines, the repulsion might not be strong enough to achieve that goal.

However, it remains unclear how those 3BFs affect the corresponding
$\Sigma$ potential.
The effective two-body forces resulting from the 3BFs considered
in Ref.~\cite{Gerstung:2020ktv} contribute not only to the $\Lambda N$ and $\Sigma N$ channels
but also to the $\Lambda N$-$\Sigma N$ transition potential.
A presently accepted constraint on the $\Sigma$ potential is
$U_\Sigma(\rho_0)=30\pm20~\text{MeV}$~\cite{Gal:2016boi},
which is inferred using data on $\Sigma^-$ atoms and on $(\pi^+,K^+)$
inclusive spectra. This constraint is fairly well met by the original chiral $YN$
potentials from 2013 and 2019 without 3BF \cite{Petschauer:2015nea,Haidenbauer:2019boi}.

We evaluate the $\Sigma$ potential in the same way as
Gerstung et al.~\cite{Gerstung:2020ktv}
have done for the $\Lambda$ potential.
The Brueckner-Hartree-Fock method with a continuous choice~\cite{Petschauer:2015nea}
is employed to calculate the hyperon single-particle potential.
Regarding the nuclear forces, the N$^3$LO $NN$ potential from Ref.~\cite{Entem:2003ft} is used,
while the $NNN$ 3BF at N$^2$LO is taken into account via a
density-dependent two-body force.
The cutoff in the regulator function~\cite{Entem:2003ft} is chosen as $500~\text{MeV}$.
The nuclear saturation properties are reproduced by the nucleon forces~\cite{Gerstung:thesis}.

\begin{figure}
	\begin{minipage}[b]{.48\linewidth}
		\centering
		\begin{tabular}{c|c}
		\hline
		$H_1~\left(f^{-2}\right)$ & $H_2~\left(f^{-2}\right)$ \\
		\hline\hline
		$-2.650$ & $0.100$ \\
		$-2.200$ & $0.000$ \\
		$-1.800$ & $-0.100$ \\
		$-1.350$ & $-0.200$ \\
		$-0.900$ & $-0.300$ \\
		\hline
		\end{tabular}
		\captionof{table}{
		Considered combinations of LECs of the $YNN$ 3BF
		that reproduce $U_\Lambda(\rho_0)=-30~\text{MeV}$ for NLO13(500).
		The values are in units of the inverse squared pion-decay constant with $f \approx 92~\text{MeV}$, and
		correspond to the left line in figure 6 of Ref.~\cite{Gerstung:2020ktv}.
		The values are taken from Gerstung's PhD thesis~\cite{Gerstung:thesis}.
		}
		\label{tab:h1h2}
	\end{minipage}
	\hfill
	\begin{minipage}[b]{.48\linewidth}
		\centering
		\includegraphics[width=\linewidth]{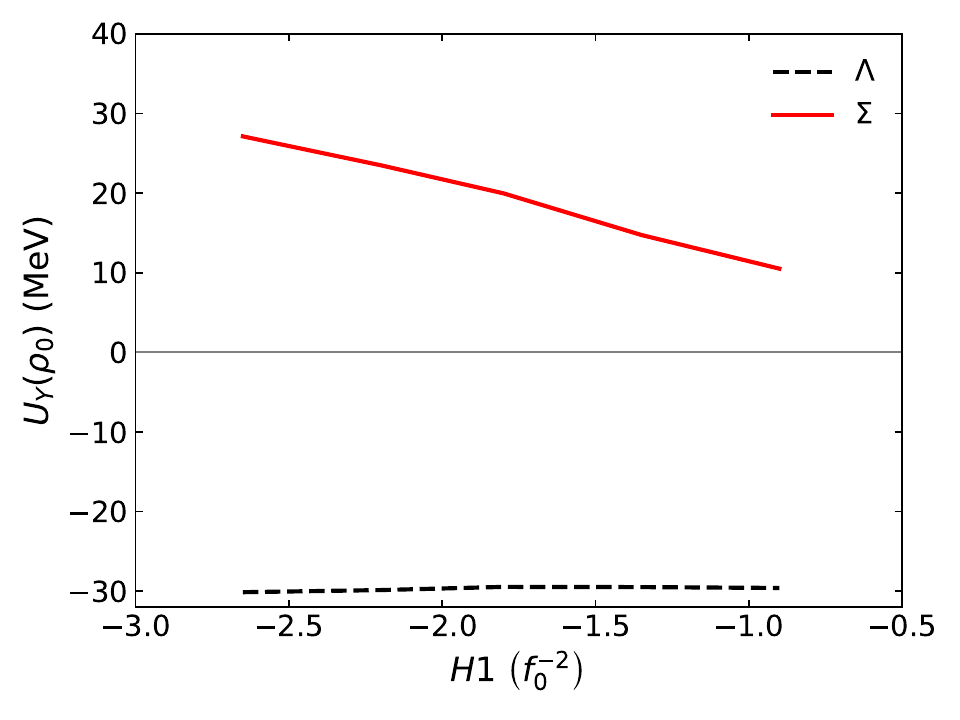}
		\captionof{figure}{$\Sigma$ (red solid) and $\Lambda$ (black dashed)
		potentials in symmetric nuclear matter. The horizontal axis corresponds
		to the three-body LECs $H_1$ with $H_2$ listed in
		table~\ref{tab:h1h2}.}
		\label{fig:UY_rho0}
	\end{minipage}\hfill
\end{figure}

\begin{figure}[thbp]
	\centering
	\includegraphics[width=0.8\hsize]{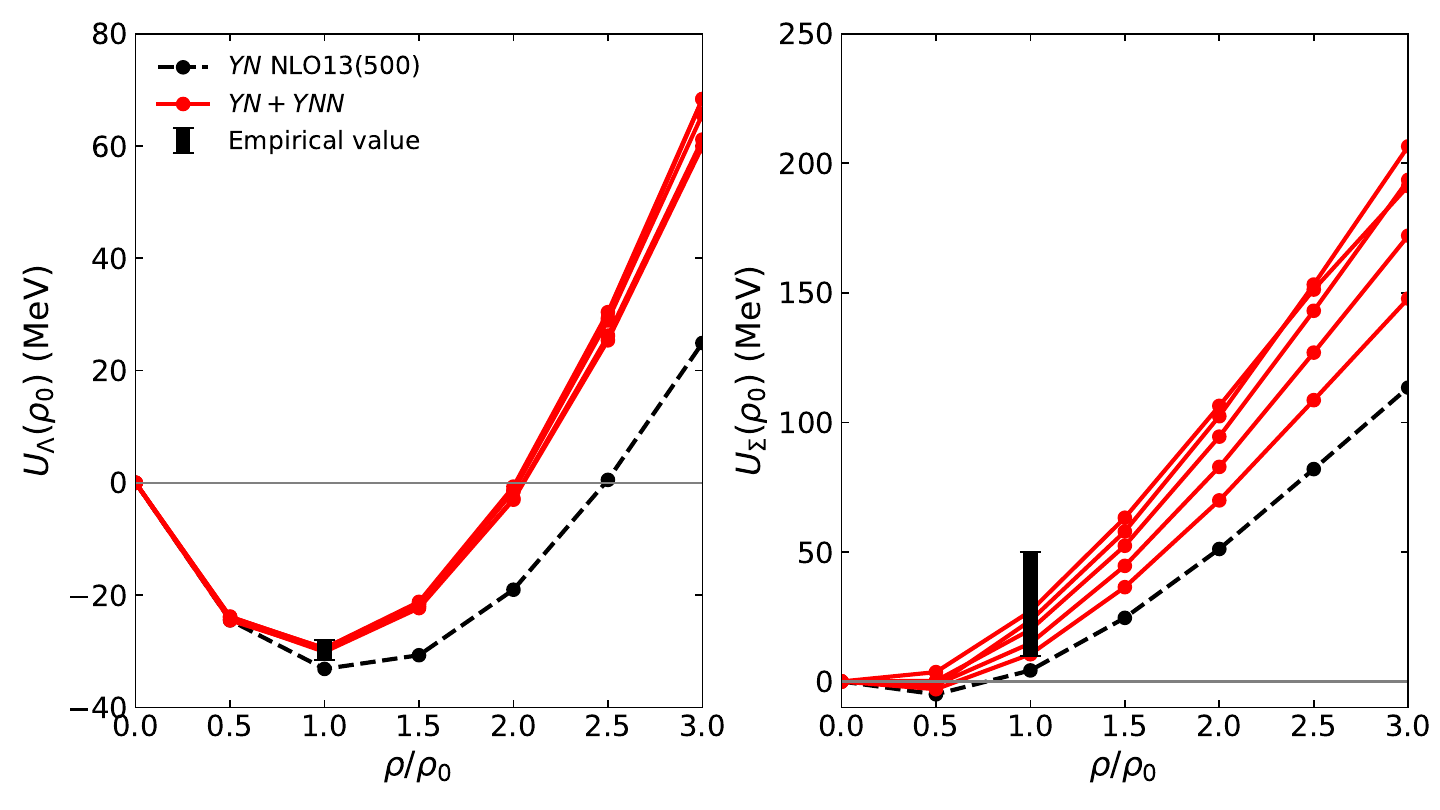}
	\caption{Density dependence of the $\Lambda$ (left panel) and $\Sigma$ (right
	panel) potentials in symmetric nuclear matter.  The red solid lines
	are calculated by using the 3BF LECs in table~\ref{tab:h1h2}.
	The only two-body case with the
	NLO13(500) parameter set is represented by the dashed line.
	The black bands show the empirical values of the $\Lambda$ potential,
	$-31.5<U_\Lambda(\rho_0)<-28.0~\text{MeV}$~\cite{Jinno:2023xjr}, and the
	$\Sigma$ potential,
	$U_\Sigma(\rho_0)=30\pm20~\text{MeV}$~\cite{Gal:2016boi}.}
	\label{fig:DensityDep_chi}
\end{figure}

For the hyperonic force, the $YN$ potential NLO13(500)~\cite{Haidenbauer:2013oca} is employed.
The $YNN$ force is implemented as an effective density-dependent $YN$ two-body force.
The number of LECs involved in the $YNN$ force is
reduced by assuming decuplet dominance approximation~\cite{Petschauer:2016pbn}.
Then, there are only three LECs: one related to
the meson-octet-decuplet baryon vertex, and two denoted by $H_1$ and $H_2$ characterizing the strengths
of the contact vertices with three-octet and one-decuplet baryons.
The meson-octet-decuplet coupling is constrained by the decay width $\Gamma(\Delta\rightarrow N \pi)$, and its large-$N_c$ value is
employed~\cite{Gerstung:2020ktv}.
The two LECs of the contact terms are fixed by requiring the reproduction of
the empirical value of the $\Lambda$ potential~\cite{Gal:2016boi,Jinno:2023xjr},
\begin{align}
	\label{eq:UL30}
	U_\Lambda(\rho_0)\simeq -30~\text{MeV},
\end{align}
inferred by using the $\Lambda$ hypernuclear spectroscopy,
and a strongly repulsive $U_\Lambda$ at high density, sufficient to resolve the
hyperon puzzle~\cite{Gerstung:2020ktv}.
Some combinations of the contact LECs that fulfill these requirements are listed
in table~\ref{tab:h1h2}.

The $\Lambda$ and $\Sigma$ potentials in symmetric nuclear matter at
$\rho_0$ are shown in figure~\ref{fig:UY_rho0} for various combinations of $H_1$ and $H_2$.
The $\Lambda$ potential is practically constant
by construction, i.e., due to the constraint~\eqref{eq:UL30}.
In contrast, the $\Sigma$ potential varies from about $30$
to $10~\text{MeV}$.

In figure~\ref{fig:DensityDep_chi}, we show the density dependence of the single-particle potentials.
One can see that certain sets of the 3BF LECs reproduce the empirical constraint
$U_\Sigma(\rho_0) = 30\pm 20~\text{MeV}$~\cite{Gal:2016boi}.
Interestingly, those 3BFs also yield $\Lambda$ potentials that are strongly repulsive
at high densities, as needed to suppress the $\Lambda$ hyperons in neutron stars.
Thus, the 3BFs can be chosen to  solve the hyperon puzzle of neutron stars while at the same time the empirical value of $U_\Sigma$ is reproduced.

\section{Summary}
\label{sec:summary}

To suppress the $\Lambda$ hyperons in neutron stars, Gerstung et
al.~\cite{Gerstung:2020ktv} calculated the $\Lambda$ single-particle potential by adding an
effective 3BF to the chiral $YN$ potentials of the J\"ulich-Bonn
group~\cite{Haidenbauer:2013oca}, which is one of the state-of-the-art modeling of
the potential based on the $\chi$EFT\@.  In this contribution, we have examined the
consistency of those potentials with the data from experiments and observations
of substantially different physics.
Specifically, we referenced the data from the $\Lambda$ hypernuclear spectroscopy,
the $\Lambda$ directed flow created in heavy-ion collisions,
and the value of the $\Sigma$ single-particle potential at the saturation density.
Some of the 3BF LEC sets of $(H_1, H_2)$ from Ref.~\cite{Gerstung:2020ktv}
turned out to be consistent with the empirical information in all three physics.

The results presented here are based on a single chiral $YN$ potential, NLO13(500).
Variants such as NLO19~\cite{Haidenbauer:2019boi} should be considered in order
to provide an estimate of the theoretical uncertainty.
Furthermore, recently a $YN$ interaction up to N$^2$LO in the chiral expansion
has been presented~\cite{Haidenbauer:2023qhf}. Initial studies suggest that it
yields more
attractive $\Lambda$ and $\Sigma$ potentials~\cite{Haidenbauer:2023qhf}.
Here, a more rigorous investigation of the in-medium properties is required.
Such a work will be performed in the future.

In addition, it is desirable to improve the theoretical treatment of the heavy-ion collision simulation.
The results shown here have been obtained by using the same $\Lambda$ potential for all other
hyperons, including their resonance states.
As a large number of $\Sigma$ hyperons and hyperon resonances are produced
during the evolution of the heavy-ion collisions, and as the behavior of the $\Lambda$
and $\Sigma$ potentials are very different, as seen in figure~\ref{fig:DensityDep_chi},
we intend to include different potentials for different hyperons to
explore their effects on $v_1$ of $\Lambda$ and $\Sigma$
in future studies.

\section{Acknowledgements}
AJ thanks Johann Haidenbauer for the collaboration and
his careful reading of the manuscript,
and Dominik Gerstung for kindly sharing his code with us.
AJ also thanks Wolfram Weise and Avraham Gal for insightful and valuable discussions
and the nuclear theory group at the Forshungszentrum J\"ulich for their splendid hospitality during
his visit where part of this work was done.
This work was supported in part by the
Grants-in-Aid for Scientific Research from JSPS
(Nos. JP21K03577 
and JP23K13102).
This work was also supported in part
by JST, the establishment of university fellowships towards
the creation of science technology innovation,
Grant No.~JPMJFS2123
and by JST SPRING, Grant Number JPMJSP2110.

\bibliographystyle{JHEP}
\bibliography{my-bib-database}

\end{document}